# Statistical inference on the $h$-index with an application to top-scientist performance

**A. Baccini[1], L. Barabesi[1], M. Marcheselli[1] and L. Pratelli[2],**

[1]Department of Economics and Statistics, University of Siena,

P.zza S.Francesco 17, 53100 Siena, Italy

[2]Naval Academy, viale Italia 72, 57100 Leghorn, Italy

**Abstract.** Despite the huge literature on h-index, few papers have been devoted to the statistical analysis of $h$-index when a probabilistic distribution is assumed for citation counts. The present contribution relies on showing the available inferential techniques, by providing the details for proper point and set estimation of the theoretical $h$-index. Moreover, some issues on simultaneous inference - aimed to produce suitable scholar comparisons - are carried out. Finally, the analysis of the citation dataset for the Nobel Laureates (in the last five years) and for the Fields medallists (from 2002 onward) is proposed .

**1. Introduction.** On August 3rd 2005, Jorge E. Hirsch uploaded an article to the arXiv.org e-Print archive (http://arxiv.org/abs/physics/0508025), in which he introduced the so-called $h$-index by means of the following definition (as given in the fifth and last uploaded version): "a scientist has index $h$ if $h$ of his/her $N_p$ papers have at least $h$ citations each, and the other $(N_p - h)$ papers have no more than $h$ citations each" (see also Hirsch, 2005). On August 18th, *Nature* - which until that time had campaigned for a moderate use of the Impact Factor - published an article (Ball, 2005), where the $h$-index was presented as "transparent, unbiased and very hard to rig" and able to "pick out influential individuals". The illustrative example in the article was a short list of seven high-ranked physicists. The top physicist was Ed

Witten of the Princeton Institute for Advanced Study "widely regarded by his peers as the most brilliant living physicist". On August 19th, a short note appeared on *Science* (Bhattacharjee, 2005), where the value of the $h$-index for Manuel Cardona, a physicist at the Max Planck Institute for Solid State Research in Stuttgart, Germany, was given. According to Bornmann and Daniel (2009), at the initiative of the same Cardona, Hirsch's paper was published a few weeks later in the *Proceedings of the National Academy of Sciences* (Hirsch, 2005).

These circumstances cannot be regarded as a matter of secondary importance in explaining the success met by the $h$-index. The immediate endorsement of the index operated by two leading scientific journals, and the publication of the paper in a prestigious outlet, were strong signals for an academic community searching for a simple and useful way to characterize the scientific output of a researcher. In addition, it was immediately realized that the $h$-index could be adopted for assessing the research performance of more complex structures, such as: journals, setting up as a competitor of the Impact Factor (Braun *et al*., 2006); groups of scholars, departments and institutions (Molinari and Molinari, 2008, van Raan, 2006) and even countries (Mahbuba and Rousseau, 2008, Nejati and Jenab, 2010). At the end of 2008, the *Web of Science* archive contained about 200 citations to the Hirsch's paper (Bornmann and Daniel, 2009) and similar figures also resulted from the *Scopus* and *Google Scholar* archives (Alonso *et al*., 2009). Two years later, the citations went up to 632 and at least 755 papers were categorized as $h$-index related literature. This literature documented the impressive diffusion of $h$-index outside the field of informetrics and scientometrics where the mainstream research is obviously developed (Zhang *et al*., 2011).

Even more impressive is the popularity of the $h$-index in non-scholarly literature. As an example, *Wired Magazine* wrote that "The $h$-index was the biggest splash in a flood of Internet-enabled rating systems - growth and decay chronometrics, semiometric measures, hub/authority metrics. Schools and labs use such ratings to help them make grants, bestow tenure, award bonuses, and hire postdocs. In fact, similar statistical approaches have become standard practice in Internet search algorithms and on social networking sites. These numbers do for scientists what […] Bill James' Sabermetrics did for baseball" (Gugliotta, 2009).

Arguably, this global success is due in large part to the simplicity of the mathematical structure of the $h$-index and its ease of calculation. A lot of papers (Alonso *et al.*, 2009, Costas and Bordons, 2007, Egghe, 2010, Rousseau, 2008) discuss advantages and disadvantages of using the $h$-index. In any case, the $h$-index is probably so diffused because it is perceived by non-technical readers as a unique numerical value measuring a very complex phenomenon such as the quality/impact/production of a researcher. Ranking scientists according to $h$-index is apparently very simple, and differences amongst researchers are directly measurable: “my $h$-index is bigger than your!”.

Scientometricians have worked on the theoretical foundations of $h$-index and three main lines of research have been explored. The first line is the deterministic one suggested by Hirsch (2005), according to which the $h$-index is the result of a linear growth model of publication and citation. More interestingly, a second line of research consists of the derivation of the $h$-index from Lotka’s law (Egghe, 2005, 2006, Egghe and Rousseau, 2006, Ye, 2011). In contrast with these mathematical model approaches, Glänzel (2006) started the third line of research, emphasizing for the first time the relevance of the “statistical background” for the $h$-index. Glänzel required that the citation number of a paper were a random variable and derived some properties of the $h$-index by assuming a Paretian model for the citation number. The importance of this approach is stressed, among others, by Rousseau (2008) and Panaretos and Malesios (2009).

When the full statistical perspective is considered, *i.e.* by assuming a statistical model for the citation-count distribution, it should be realized that the original definition provided by Jorge E. Hirsch gives rise to an empirical index and that the corresponding theoretical index has to be properly defined. Obviously, this process is - in some way - statistically unsound, since the “estimator” is defined in advance to the “parameter” to be estimated. In any case, once that the definition of the theoretical index is suitably carried out, the statistical properties of the empirical index must be assessed. Even if Glänzel (2006) produced the first effort in this direction, the decisive step was made by Beirlant and Einmahl (2010) who handled the empirical $h$-index as the estimator of a suitable statistical functional of the citation-count distribution. Beirlant and Einmahl (2010) also gave the consistency of the empirical $h$-index with respect to this functional and the conditions for its large-sample normality. In addition, they provided a variance estimation procedure when the underlying citation-count distribution displays Pareto-type or Weibull-type tails. However, Beirlant and Einmahl (2010) stated their theory by assuming a continuous citation-count distribution, even if the citation number is obviously an integer. Hence, Pratelli *et al.* (2012) further developed the results by Beirlant and Einmahl (2010), by achieving similar findings when the citation count follows a distribution supported by the integers. In addition, Pratelli *et al.* (2012) provided a suitable expression for the variance of empirical $h$-index, which allows for simple and consistent nonparametric variance estimation. In turn, on the basis of these results, large-sample nonparametric confidence intervals may be implemented.

Panaretos and Malesios (2009) remarked that “while there exists a vast literature on the empirical $h$-index and its applications, relatively little work has been done on the study of the theoretical $h$-index as a statistical function, allowing to construct confidence intervals, test hypotheses and check the validity of its statistical properties”. Hence, the aim of the present paper is to divulge the available statistical tools for the inference on the $h$-index, by trying to explain issues and details which may be obscure for non-statisticians. Moreover, an extensive application to real data is given, in order to highlight the importance of producing interval estimation, in addition to point estimation. Finally, simultaneous inference techniques are introduced in order to achieve suitable scholar comparisons.

**2. The empirical and theoretical $h$-index.** Let us assume that $X$ be an integer-valued random variable representing the citation number for a paper of a given scholar. Moreover, it is assumed that $S$ be the survival function corresponding to the random variable $X$, *i.e.* $S(x) = P(X > x)$. Therefore, $S(x)$ constitutes the probability that a paper of the scholar receives more than $x$ citation. The random variable $X$ is usually required to be “heavy-tailed” in the scientometric applications (see *e.g.* Glänzel, 2006, 2010), even if the results given in this section hold in general. Hence, if the scholar has published $n$ papers, the random variables $X_1, \ldots, X_n$ represent the citation counts for his/her $n$ papers. In order to develop the theory, it is required that $X_1, \ldots, X_n$ be identically and independently distributed.

On the basis of the Hirsch's definition given in the Introduction, the empirical $h$-index - say $\widehat{H}$ - may be mathematically expressed as

$$\widehat{H} = \max\{j \in \mathbb{N} : n\widehat{S}(j-1) \geq j\} , \qquad (1)$$

where $\widehat{S}$ represents the empirical survival function, *i.e.*

$$\widehat{S}(x) = \frac{1}{n}\sum_{i=1}^{n} I_{(x,\infty)}(X_i) ,$$

while $I_E$ turns out to be the usual indicator function of a set $E$, *i.e.* $I_E(x) = 1$ if $x \in E$ and $I_E(x) = 0$ otherwise. Obviously, $\widehat{S}(x)$ is the empirical rate of citation counts greater than a given $x$ and hence $\widehat{S}$ is the "natural" estimator of $S$. Moreover, it is apparent that the quantity $n\widehat{S}(j-1)$ represents the number of papers receiving at least $j$ citations. Thus, it is immediate to realize that expression (1) formally states the empirical $h$-index in accordance with the definition provided by Hirsch (2005). However, Pratelli *et al.* (2012) emphasized that (1) gives rise to the following alternative and equivalent (but more convenient) expression for $\widehat{H}$, *i.e.*

$$\widehat{H} = \sum_{j=1}^{n} I_{[j/n,1]}(\widehat{S}(j-1)) . \qquad (2)$$

Obviously, $\widehat{H}$ is a random variable since it depends on the random variables $X_1, \ldots, X_n$. Moreover, it should be noticed from (2) that $\widehat{H} = f(\widehat{S})$, *i.e.* the empirical $h$-index is actually a functional of the empirical survival function. This remark allows for a suitable definition of the theoretical $h$-index - say $h$ - which may be inherently defined as $h = f(S)$ by adopting the statistical "correspondence principle". More precisely, on the basis of expression (2) the theoretical $h$-index may be set to

$$h = \sum_{j=1}^{n} I_{[j/n,1]}(S(j-1)) , \qquad (3)$$

as suggested by Pratelli *et al.* (2012). Obviously, $h$ depends on $n$ and it is easily verified that $h \to \infty$ and $h/n \to 0$ as $n \to \infty$, a quite unusual behavior for a statistical parameter.

As to the main statistical properties of the empirical $h$-index, Pratelli *et al.* (2012) proved that

$$\mathrm{E}[\widehat{H}] = \sum_{j=1}^{n} p_j$$

and

$$\mathrm{Var}[\widehat{H}] = \sum_{j=1}^{n} p_j(1-p_j) + 2\sum_{l=2}^{n}\sum_{j=1}^{l-1} p_l(1-p_j) \ ,$$

where

$$p_j = \sum_{l=j}^{n} \binom{n}{l} S(j-1)^l (1-S(j-1))^{n-l} \ .$$

Thus, it is apparent that $\widehat{H}$ is a biased estimator for $h$. However, since Pratelli *et al.* (2012) showed that

$$\lim_n \mathrm{E}\left[\left(\frac{\widehat{H}}{h} - 1\right)^2\right] = 0 \ ,$$

it also follows that $\widehat{H}/h \xrightarrow{P} 1$ as $n \to \infty$, *i.e.* the ratio $\widehat{H}/h$ converges in probability to one. Thus, $\widehat{H}$ may be considered as a "consistent" estimator for $h$, even if in this setting the usual definition of consistency is pointless since the parameter approaches to infinity as simple size increases (see also a similar comment by Beirlant and Einmahl, 2010).

As previously emphasized, in the present framework some arbitrariness arises in the choice of the theoretical $h$-index and hence some attention should be put in order to properly identify the reference parameter under estimation. Since in many statistical applications the expected value of the estimator coincides with the parameter to be estimated, we argue that $\mathrm{E}[\widehat{H}]$ could be considered as a "natural" competitor of $h$. This suggestion is also supported by the equivalence of $h$ and $\mathrm{E}[\widehat{H}]$ for large $n$, *i.e.*

$$\lim_n \frac{\mathrm{E}[\widehat{H}]}{h} = 1 \ ,$$

as shown by Pratelli *et al.* (2012). It is worth noting that $h$ solely assumes integer values, while $\mathrm{E}[\widehat{H}]$ is real-valued. In any case, many simulation studies carried out by Pratelli *et al.* (2012) have shown that $h$ and $\mathrm{E}[\widehat{H}]$ are very similar even for small $n$.

**3. Large-sample properties of the empirical $h$-index.** With the aim of achieving the implementation of large-sample confidence intervals for $h$ or $\mathrm{E}[\widehat{H}]$, the assessment of the large-sample properties of $\mathrm{Var}[\widehat{H}]$ is of central importance. It is woth noting that, since $\mathrm{E}[\widehat{H}] \to \infty$ as $n \to \infty$ and since scientometricians usually required "heavy-tailed" distributions for the citation counts, the most interesting case should imply that $\mathrm{Var}[\widehat{H}] \to \infty$ as $n \to \infty$.

First, in order to obtain a conservative estimator of $\mathrm{Var}[\widehat{H}]$, it is useful to introduce an operative condition on the underlying citation distribution, *i.e.* for each $M > 0$ it is assumed that

$$\lim_n \left( \sup_{j \in D_M} \left| \frac{P(X=j)}{P(X=n)} - 1 \right| \right) = 0\,, \tag{4}$$

where $D_M = [n - M\sqrt{n}, n + M\sqrt{n}] \cap \mathbb{N}$. Intuitively, if the random variable $X$ satisfies the condition (4), for a large $n$ its distribution is nearly uniform on an interval of natural numbers centered on $n$ and with size proportional to $\sqrt{n}$. Practically speaking, assumption (4) actually implies a "slow decrement" of $P(X = n)$ as $n \to \infty$. Actually, the underlying citation distribution is commonly assumed to be Pareto-type or Weibull-type (see *e.g.* Glänzel, 2006, 2010, Barcza and Telcs, 2009, Beirlant and Einmahl, 2010) and these distribution types - or their mixtures - satisfy condition (4). As a matter of fact, if $l$ is a slowly-varying function, *i.e.* $l(tx)/l(t) \to 1$ for each $x$ as $t \to \infty$, a Pareto-type distribution is characterized by a survival function given by $S(x) = x^{-\alpha} l(x)$ and hence it verifies (4) for any $\alpha > 0$. This distribution type encompasses families of central importance for describing heavy-tailed discrete data, such as the discrete stable distribution (see *e.g.* Marcheselli *et al.*, 2008). Analogously, a Weibull-type distribution is characterized by a survival function given by $S(x) = \exp(-x^{\tau} l(x))$ and accordingly it verifies (4) for any $\tau < 1/2$.

A "natural" estimator for the quantity $p_j$ may be obtained in by means of a plug-in of the empirical survival function into the expression of $p_j$, *i.e.*

$$\widehat{p}_j = \sum_{l=j}^{n} \binom{n}{l} \widehat{S}(j-1)^l (1-\widehat{S}(j-1))^{n-l} \, .$$

Hence, on the basis of the expression of $\mathrm{Var}[\widehat{H}]$, by adopting in turn the statistical "correspondence principle", Pratelli *et al.* (2012) propose the variance estimator

$$\widehat{V} = \sum_{j=1}^{\min(\lfloor 3\widehat{H} \rfloor, n)} \widehat{p}_j(1-\widehat{p}_j) + 2 \sum_{l=2}^{\min(\lfloor 3\widehat{H} \rfloor, n)} \sum_{j=1}^{l-1} \widehat{p}_l \, (1-\widehat{p}_j) \, , \tag{5}$$

where $\lfloor x \rfloor$ denotes the greatest integer less than or equal to $x$. In expression (5), the truncation of the summation extremes is due to some technical issues in order to improve estimation (see Pratelli *et al.*, 2012). Under condition (4) it generally holds that

$$\lim_{n} \mathrm{Var}[\widehat{H}] = \infty \, ,$$

and Pratelli *et al.* (2012) proved the "consistency" of the estimator $\widehat{V}$, in the sense that

$$\frac{\widehat{V}}{\mathrm{Var}[\widehat{H}]} \xrightarrow{P} 1$$

as $n \to \infty$. In turn, equivalently to the Section 2, the usual definition of consistency is not useful since $\mathrm{Var}[\widehat{H}]$ approaches to infinity as simple size increases. It should be also remarked that estimator (5) is fully nonparametric since it does not require the specification of a model for the underlying citation distribution. In contrast, the variance estimator proposed by Beirlant and Einmahl (2010) assumes semi-parametric modelling and it implies the estimation of the Paretian index for the Pareto-type family, which is a complicate task. In addition, the computation of estimator (5) is straightforward from a practical point of view.

On the basis of the findings by Pratelli *et al.* (2012), if condition (4) is verified, the large-sample normality of $\widehat{H}$ holds, *i.e.* the following convergences in distribution are achieved

$$\frac{\widehat{H}-h}{\sqrt{\widehat{V}}} \sim \frac{\widehat{H}-\mathrm{E}[\widehat{H}]}{\sqrt{\widehat{V}}} \xrightarrow{d} N(0,1) \, ,$$

as $n \to \infty$, where - as usual - $N(0,1)$ represents a standard Normal random variable. The previous result provides the pivotal quantities for the implementation of a large-sample confidence set at the $(1-\gamma)$ confidence level for $h$ given by

$$C = \{[\![\widehat{H} - z_{1-\gamma/2}\sqrt{\widehat{V}}]\!], \ldots, [\![\widehat{H} + z_{1-\gamma/2}\sqrt{\widehat{V}}]\!]\}\ ,$$

where $z_\gamma$ represents the $\gamma$-th quantile of the standard Normal distribution, while $[\![x]\!]$ represents the integer closest to $x$. Obviously, $C$ turns out to be a confidence set since $h$ may solely assume integer values. Similarly, a large-sample confidence interval at the $(1-\gamma)$ confidence level for $\mathrm{E}[\widehat{H}]$ is given by

$$C' = (\widehat{H} - z_{1-\gamma/2}\sqrt{\widehat{V}}, \widehat{H} + z_{1-\gamma/2}\sqrt{\widehat{V}})\ .$$

It should be again emphasized that $C$ and $C'$ are fully nonparametric confidence set and interval, respectively. Indeed, their implementation does not demand the specification of a distribution and solely requires the validity of condition (4), which is likely to hold for almost all the distributions of interest in the area of scientometrics. Moreover, a large simulation study carried out by Pratelli *et al.* (2012) show that the actual coverage of $C$ and $C'$ is appropriate even for quite small $n$.

In the case that $k$ scholars have to be jointly compared, a suitable procedure should be applied in order to achieve simultaneous inference. Let us suppose that the $k$ scholars act independently and that they have published $n_1, \ldots, n_k$ papers each, while their corresponding $k$ theoretical $h$-index are given by $h_1, \ldots, h_k$. Accordingly, let $\widehat{H}_1, \ldots, \widehat{H}_k$ be the empirical $h$-indexes of these scholars and let $\widehat{V}_1, \ldots, \widehat{V}_k$ be the variance estimators. Thus, on the basis of the procedure suggested by Šidák (1967), $k^* = k(k-1)/2$ large-sample conservative simultaneous confidence sets for the differences $(h_j - h_l)$ $(l > j = 1, \ldots, k)$ at the $(1-\gamma)$ confidence level are given by

$$C_{jl} = \{[\![\widehat{H}_j - \widehat{H}_l - z_{\gamma^*}\sqrt{\widehat{V}_{jl}}]\!], \ldots, [\![\widehat{H}_j - \widehat{H}_l + z_{\gamma^*}\sqrt{\widehat{V}_{jl}}]\!]\}\ ,$$

where $\widehat{V}_{jl} = \widehat{V}_j + \widehat{V}_l$, while $\gamma^* = (1 + (1-\gamma)^{1/k^*})/2$. Similarly, $k^*$ large-sample conservative simultaneous confidence sets for the differences $(\mathrm{E}[\widehat{H}_j] - \mathrm{E}[\widehat{H}_l])$ $(l > j = 1, \ldots, k)$ at the $(1-\gamma)$ confidence level are given by

$$C'_{jl} = (\widehat{H}_j - \widehat{H}_l - z_{\gamma^*}\sqrt{\widehat{V}_{jl}}, \widehat{H}_j - \widehat{H}_l + z_{\gamma^*}\sqrt{\widehat{V}_{jl}})\ .$$

Obviously, more refined simultaneous procedure could be implemented, such as the bootstrap techniques recently suggested by Mandel and Betensky (2008) or by Xiong (2011).

**4. Some real data examples.** In order to exemplify the discussed statistical tools, we have considered the citation datasets of the Nobel Laureates in the last five years and of the Fields medallists from 2002 onward. Citation performances of these authors are drawn from an author search on *Scopus* carried out during February 2012. Tables I and II present the analyzed scholars, who are ordered according to their empirical $h$-indexes for each discipline. Moreover, in these Tables the number of papers, the empirical $h$-index and the large-sample confidence set at the $95\%$ confidence level are given for each scholar.

As a specific example for the statistical interpretation of Tables I and II, Adrian Fert - winner of the Nobel Prize for Physics in 2007 - displays an empirical $h$-index equal to $52$, which is the highest for physicists. Once that the inferential paradigm is assumed, $\widehat{H} = 52$ constitutes a point estimate of the (unknown) theoretical $h$-index and it should be coupled with an estimate of the sampling variability. Loosely speaking, the set estimate - *i.e.* the corresponding confidence set $\{46, \ldots, 58\}$ - allows for jointly assessing the two aspects. The point estimate and the confidence set of Adrian Fert are not really different from those of Andre Geim, the second in this ranking. In contrast, Konstantin Novoselov displays an empirical $h$-index equal to $42$, even if the corresponding confidence set, *i.e.* $\{35, \ldots, 49\}$, overlaps the confidence sets of the previous physicists.

**Tables I and II about here**

A simple analysis of the Tables I and II leads to three main conclusions. The first argument is well-known, in the sense that top scientists in different disciplines have different scientometric indexes. These differences mainly depend on the specific pattern of productivity and on the citation habits of the discipline. The second conclusion relies on the fact that in each discipline the use of the $h$-index flattens the performance of scholars. The choice of a unique value - synthesizing the individual productivity and the citations received

- tends to equalize very different publication behaviors adopted by different scholars. As an example, in Physics Adrian Fert and Brian Schmidt have similar empirical $h$-indexes, even if Adrian Fert published a number of papers which is double with respect to Brian Schmidt. The third conclusion is the most striking one: in each discipline the majority of confidence sets intersects, so that a strict ranking of the considered scholars may not be feasible. This is a very important issue, since the common use of the $h$-index aims to rank individuals, journals and so on. If these rankings fail to consider the sample variability, the differences between scholars in different positions may be not more than an optical illusion.

In order to show the practical implementation of simultaneous inference, the Nobel Laureates for Economics have been analyzed. Since in this group there are $k = 10$ scholars with theoretical $h$-indexes given by $h_1, \ldots, h_{10}$, $k^* = 45$ differences $(h_j - h_l)$ $(l > j = 1, \ldots, 10)$ must be considered. Table III reports the corresponding large-sample pairwise simultaneous confidence sets at the $95\%$ confidence level. By analyzing Table III, if the simultaneous confidence sets not containing the zero are considered, some orderings on the theoretical $h$-indexes may be statistically stated. More precisely, by considering the first nine confidence intervals, it follows that $h_1 > h_8, h_9, h_{10}$; the subsequent eight confidence intervals provide $h_2 > h_8, h_9, h_{10}$; the subsequent seven confidence intervals provide $h_3 > h_9, h_{10}$; the subsequent six confidence intervals provide $h_4 > h_{10}$; the subsequent five confidence intervals provide $h_5 > h_{10}$; the subsequent four confidence intervals provide $h_6 > h_{10}$; the subsequent three confidence intervals provide $h_7 > h_{10}$. Hence, a strict statistical ranking of these scholars is not available. Indeed, in synthesis, it can be solely stated that $h_1, h_2 > h_8, h_9, h_{10}$, $h_3 > h_9, h_{10}$ and $h_4, h_5, h_6, h_7 > h_{10}$ at the $95\%$ confidence level.

**Table III about here**

The previous analyses emphasize that the application of the correct statistical approach should be demanded in bibliometrics and scientometrics, where the adopted methods often appear pre-statistical and pre-inferential. As noticed by Peter Hall "... issues that are obvious to statisticians are often ignored in bibliometric analysis ...”, and for example “... many proponents of impact factors, and other aspects of citation analysis, have little concept of the problems caused by averaging very heavy tailed data ...” (IMS Presidential Address, *IMS Bulletin Online*, September 2, 2011). On the other hand, Peter Hall concludes that “... we should definitely take a greater interest in this area”. Indeed, also in our opinion, scientometricians and statisticians should be more and more cooperative in order to achieve a proper development for the evaluation of the scientific performance.

**Table I.** Citation performance of the considered Nobel winners.

| $j$ | Physics | $n_j$ | $\widehat{H}_j$ | $C_j$ | Medicine | $n_j$ | $\widehat{H}_j$ | $C_j$ | Chemistry | $n_j$ | $\widehat{H}_j$ | $C_j$ | Economics | $n_j$ | $\widehat{H}_j$ | $C_j$ |
|---|---|---|---|---|---|---|---|---|---|---|---|---|---|---|---|---|
| 1 | Fert, A. (2007) | 254 | 52 | $\{46,\dots,58\}$ | Steinmann, R.M. (2011) | 412 | 110 | $\{102,\dots,118\}$ | Tsien, R.Y. (2008) | 268 | 101 | $\{92,\dots,110\}$ | Ostrom, E. (2009) | 97 | 29 | $\{24,\dots,34\}$ |
| 2 | Geim, A. (2010) | 196 | 51 | $\{43,\dots,59\}$ | Beutler, B.A. (2011) | 308 | 77 | $\{69,\dots,85\}$ | Steitz, T.A. (2009) | 247 | 79 | $\{69,\dots,89\}$ | Krugman, P. (2008) | 71 | 29 | $\{22,\dots,36\}$ |
| 3 | Schmidt, B.P. (2011) | 126 | 45 | $\{39,\dots,51\}$ | Hoffmann, J.A. (2011) | 203 | 71 | $\{61,\dots,81\}$ | Ertl, G. (2007) | 545 | 72 | $\{66,\dots,78\}$ | Sargent, T. (2011) | 85 | 25 | $\{21,\dots,29\}$ |
| 4 | Novoselov, K. (2010) | 133 | 42 | $\{35,\dots,49\}$ | Smithies, O. (2007) | 297 | 66 | $\{57,\dots,75\}$ | Negishi, E. (2010) | 316 | 53 | $\{48,\dots,58\}$ | Diamond, P.A. (2010) | 49 | 19 | $\{15,\dots,23\}$ |
| 5 | Perlmutter, S. (2011) | 133 | 38 | $\{32,\dots,44\}$ | Capecchi, M. (2007) | 190 | 62 | $\{54,\dots,70\}$ | Ramakrishnan, V.R. (2009) | 100 | 47 | $\{41,\dots,53\}$ | Maskin, E.S. (2007) | 55 | 19 | $\{14,\dots,24\}$ |
| 6 | Riess, A. (2011) | 95 | 36 | $\{30,\dots,42\}$ | Blackburn, E.H. (2009) | 227 | 59 | $\{51,\dots,67\}$ | Suzuki, A. (2010) | 768 | 39 | $\{33,\dots,45\}$ | Myerson, R.B. (2007) | 47 | 19 | $\{15,\dots,23\}$ |
| 7 | Kobayashi, M. (2008) | 386 | 35 | $\{32,\dots,38\}$ | Greider, C.W (2009) | 97 | 58 | $\{51,\dots,65\}$ | Shimomura, O. (2008) | 259 | 39 | $\{35,\dots,43\}$ | Pissarides, C.A. (2010) | 37 | 17 | $\{12,\dots,22\}$ |
| 8 | Grunberg, P.A. (2007) | 132 | 21 | $\{18,\dots,24\}$ | Szostak, J.W. (2009) | 203 | 56 | $\{48,\dots,64\}$ | Chalfie, M. (2008) | 102 | 39 | $\{34,\dots,44\}$ | Sims, C.A. (2011) | 36 | 15 | $\{10,\dots,20\}$ |
| 9 | Nambu, Y. (2008) | 80 | 17 | $\{13,\dots,21\}$ | zur Hauser, H. (2008) | 338 | 54 | $\{48,\dots,60\}$ | Heck, R.F. (2010) | 106 | 30 | $\{24,\dots,36\}$ | Mortensen, D.A. (2010) | 28 | 12 | $\{8,\dots,16\}$ |
| 10 | Boyle, W.S. (2009) | 23 | 7 | $\{4,\dots,10\}$ | Barré-Sinoussi, F. (2008) | 251 | 45 | $\{40,\dots,50\}$ | Yonath, A. (2009) | 145 | 30 | $\{25,\dots,35\}$ | Hurwicz, L. (2007) | 20 | 7 | $\{5,\dots,9\}$ |
| 11 | Smith, G.E. (2009) | 30 | 5 | $\{3,\dots,7\}$ | Evans, M.J. (2007) | 141 | 44 | $\{37,\dots,51\}$ | Shechtman, D. (2011) | 81 | 16 | $\{13,\dots,19\}$ | | | | |
| 12 | Kao, C.K. (2009) | 20 | 1 | $\{0,\dots,2\}$ | Montagnier, L. (2008) | 311 | 42 | $\{36,\dots,48\}$ | | | | | | | | |
| 13 | Maskawa, T. (2008) | 5 | 1 | $\{0,\dots,3\}$ | Edwards, R.G. (2010) | 316 | 41 | $\{35,\dots,47\}$ | | | | | | | | |

**Table II.** Citation performance of the considered Field medallists.

| $j$ | Field medallists | $n_j$ | $\widehat{H}_j$ | $C_j$ |
|---|---|---|---|---|
| 1 | Tao, T. (2006) | 164 | 29 | $\{25,\dots,33\}$ |
| 2 | Villani, C. (2010) | 55 | 21 | $\{16,\dots,26\}$ |
| 3 | Okounkov, A. (2006) | 48 | 18 | $\{16,\dots,20\}$ |
| 4 | Werner, W. (2006) | 39 | 16 | $\{12,\dots,20\}$ |
| 5 | Lindenstrauss, E. (2010) | 26 | 8 | $\{5,\dots,11\}$ |
| 6 | Smirnov, S. (2010) | 24 | 8 | $\{6,\dots,10\}$ |
| 7 | Bao Chau, N. (2010) | 9 | 7 | $\{5,\dots,9\}$ |
| 8 | Voevodsky, V. (2002) | 12 | 6 | $\{3,\dots,9\}$ |
| 9 | Lafforgue, L. (2002) | 5 | 2 | $\{0,\dots,4\}$ |
| 10 | Perelman, G. (2006) | 2 | 1 | $\{0,\dots,2\}$ |

**Table III.** Pairwise simultaneous confidence sets of the Nobel Laureates for Economics.

| $h_j - h_l$ | $C_{jl}$ | $h_j - h_l$ | $C_{jl}$ |
|---|---|---|---|
| $h_1 - h_2$ | $\{-14, \dots, 14\}$ | $h_4 - h_5$ | $\{-11, \dots, 11\}$ |
| $h_1 - h_3$ | $\{-7, \dots, 15\}$ | $h_4 - h_6$ | $\{-10, \dots, 10\}$ |
| $h_1 - h_4$ | $\{-1, \dots, 21\}$ | $h_4 - h_7$ | $\{-9, \dots, 13\}$ |
| $h_1 - h_5$ | $\{-2, \dots, 22\}$ | $h_4 - h_8$ | $\{-6, \dots, 14\}$ |
| $h_1 - h_6$ | $\{-1, \dots, 21\}$ | $h_4 - h_9$ | $\{-3, \dots, 17\}$ |
| $h_1 - h_7$ | $\{0, \dots, 24\}$ | $h_4 - h_{10}$ | $\{4, \dots, 20\}$ |
| $h_1 - h_8$ | $\{2, \dots, 26\}$ | | |
| $h_1 - h_9$ | $\{6, \dots, 28\}$ | $h_5 - h_6$ | $\{-10, \dots, 10\}$ |
| $h_1 - h_{10}$ | $\{13, \dots, 31\}$ | $h_5 - h_7$ | $\{-9, \dots, 13\}$ |
| | | $h_5 - h_8$ | $\{-7, \dots, 15\}$ |
| $h_2 - h_3$ | $\{-9, \dots, 17\}$ | $h_5 - h_9$ | $\{-4, \dots, 18\}$ |
| $h_2 - h_4$ | $\{-3, \dots, 23\}$ | $h_5 - h_{10}$ | $\{3, \dots, 21\}$ |
| $h_2 - h_5$ | $\{-4, \dots, 24\}$ | | |
| $h_2 - h_6$ | $\{-3, \dots, 23\}$ | $h_6 - h_7$ | $\{-8, \dots, 12\}$ |
| $h_2 - h_7$ | $\{-2, \dots, 26\}$ | $h_6 - h_8$ | $\{-6, \dots, 14\}$ |
| $h_2 - h_8$ | $\{1, \dots, 27\}$ | $h_6 - h_9$ | $\{-3, \dots, 17\}$ |
| $h_2 - h_9$ | $\{4, \dots, 30\}$ | $h_6 - h_{10}$ | $\{4, \dots, 20\}$ |
| $h_2 - h_{10}$ | $\{10, \dots, 34\}$ | | |
| | | $h_7 - h_8$ | $\{-9, \dots, 13\}$ |
| $h_3 - h_4$ | $\{-4, \dots, 16\}$ | $h_7 - h_9$ | $\{-6, \dots, 16\}$ |
| $h_3 - h_5$ | $\{-4, \dots, 16\}$ | $h_7 - h_{10}$ | $\{2, \dots, 18\}$ |
| $h_3 - h_6$ | $\{-3, \dots, 15\}$ | | |
| $h_3 - h_7$ | $\{-2, \dots, 18\}$ | $h_8 - h_9$ | $\{-7, \dots, 13\}$ |
| $h_3 - h_8$ | $\{0, \dots, 20\}$ | $h_8 - h_{10}$ | $\{0, \dots, 16\}$ |
| $h_3 - h_9$ | $\{3, \dots, 23\}$ | | |
| $h_3 - h_{10}$ | $\{11, \dots, 25\}$ | $h_9 - h_{10}$ | $\{-3, \dots, 13\}$ |